\documentclass[
twocolumn,
superscriptaddress,
showpacs,preprintnumbers,
amsmath,amssymb,
aps,
prl,
longbibliography
]{revtex4-1}

\usepackage{tabularx}
\usepackage{multirow}
\usepackage{xcolor}
\usepackage{graphicx}
\usepackage{subfigure}
\usepackage{dcolumn}
\usepackage{bm}
\usepackage{notes2bib} 
\usepackage[colorlinks = true,
            linkcolor = red,
            urlcolor  = blue,
            citecolor = red,
            anchorcolor = blue]{hyperref}
\usepackage{lineno}

\begin{document}

\title{First Measurement of Inclusive Electron-Neutrino and Antineutrino Charged Current Differential Cross Sections in Charged Lepton Energy on Argon in MicroBooNE}

\newcommand{\Bern}{Universit{\"a}t Bern, Bern CH-3012, Switzerland}
\newcommand{\BNL}{Brookhaven National Laboratory (BNL), Upton, NY, 11973, USA}
\newcommand{\UCSB}{University of California, Santa Barbara, CA, 93106, USA}
\newcommand{\Cambridge}{University of Cambridge, Cambridge CB3 0HE, United Kingdom}
\newcommand{\CIEMAT}{Centro de Investigaciones Energ\'{e}ticas, Medioambientales y Tecnol\'{o}gicas (CIEMAT), Madrid E-28040, Spain}
\newcommand{\Chicago}{University of Chicago, Chicago, IL, 60637, USA}
\newcommand{\Cincinnati}{University of Cincinnati, Cincinnati, OH, 45221, USA}
\newcommand{\CSU}{Colorado State University, Fort Collins, CO, 80523, USA}
\newcommand{\Columbia}{Columbia University, New York, NY, 10027, USA}
\newcommand{\Edinburgh}{University of Edinburgh, Edinburgh EH9 3FD, United Kingdom}
\newcommand{\FNAL}{Fermi National Accelerator Laboratory (FNAL), Batavia, IL 60510, USA}
\newcommand{\Granada}{Universidad de Granada, Granada E-18071, Spain}
\newcommand{\Harvard}{Harvard University, Cambridge, MA 02138, USA}
\newcommand{\IIT}{Illinois Institute of Technology (IIT), Chicago, IL 60616, USA}
\newcommand{\KSU}{Kansas State University (KSU), Manhattan, KS, 66506, USA}
\newcommand{\Lancaster}{Lancaster University, Lancaster LA1 4YW, United Kingdom}
\newcommand{\LANL}{Los Alamos National Laboratory (LANL), Los Alamos, NM, 87545, USA}
\newcommand{\Manchester}{The University of Manchester, Manchester M13 9PL, United Kingdom}
\newcommand{\MIT}{Massachusetts Institute of Technology (MIT), Cambridge, MA, 02139, USA}
\newcommand{\Michigan}{University of Michigan, Ann Arbor, MI, 48109, USA}
\newcommand{\Minnesota}{University of Minnesota, Minneapolis, MN, 55455, USA}
\newcommand{\NMSU}{New Mexico State University (NMSU), Las Cruces, NM, 88003, USA}
\newcommand{\Oxford}{University of Oxford, Oxford OX1 3RH, United Kingdom}
\newcommand{\Pitt}{University of Pittsburgh, Pittsburgh, PA, 15260, USA}
\newcommand{\Rutgers}{Rutgers University, Piscataway, NJ, 08854, USA}
\newcommand{\SLAC}{SLAC National Accelerator Laboratory, Menlo Park, CA, 94025, USA}
\newcommand{\SDSMT}{South Dakota School of Mines and Technology (SDSMT), Rapid City, SD, 57701, USA}
\newcommand{\Maine}{University of Southern Maine, Portland, ME, 04104, USA}
\newcommand{\Syracuse}{Syracuse University, Syracuse, NY, 13244, USA}
\newcommand{\TelAviv}{Tel Aviv University, Tel Aviv, Israel, 69978}
\newcommand{\Tennessee}{University of Tennessee, Knoxville, TN, 37996, USA}
\newcommand{\UTA}{University of Texas, Arlington, TX, 76019, USA}
\newcommand{\Tufts}{Tufts University, Medford, MA, 02155, USA}
\newcommand{\VTech}{Center for Neutrino Physics, Virginia Tech, Blacksburg, VA, 24061, USA}
\newcommand{\Warwick}{University of Warwick, Coventry CV4 7AL, United Kingdom}
\newcommand{\Yale}{Wright Laboratory, Department of Physics, Yale University, New Haven, CT, 06520, USA}

\affiliation{\Bern}
\affiliation{\BNL}
\affiliation{\UCSB}
\affiliation{\Cambridge}
\affiliation{\CIEMAT}
\affiliation{\Chicago}
\affiliation{\Cincinnati}
\affiliation{\CSU}
\affiliation{\Columbia}
\affiliation{\Edinburgh}
\affiliation{\FNAL}
\affiliation{\Granada}
\affiliation{\Harvard}
\affiliation{\IIT}
\affiliation{\KSU}
\affiliation{\Lancaster}
\affiliation{\LANL}
\affiliation{\Manchester}
\affiliation{\MIT}
\affiliation{\Michigan}
\affiliation{\Minnesota}
\affiliation{\NMSU}
\affiliation{\Oxford}
\affiliation{\Pitt}
\affiliation{\Rutgers}
\affiliation{\SLAC}
\affiliation{\SDSMT}
\affiliation{\Maine}
\affiliation{\Syracuse}
\affiliation{\TelAviv}
\affiliation{\Tennessee}
\affiliation{\UTA}
\affiliation{\Tufts}
\affiliation{\VTech}
\affiliation{\Warwick}
\affiliation{\Yale}

\author{P.~Abratenko} \affiliation{\Tufts} 
\author{R.~An} \affiliation{\IIT}
\author{J.~Anthony} \affiliation{\Cambridge}
\author{L.~Arellano} \affiliation{\Manchester}
\author{J.~Asaadi} \affiliation{\UTA}
\author{A.~Ashkenazi}\affiliation{\TelAviv}
\author{S.~Balasubramanian}\affiliation{\FNAL}
\author{B.~Baller} \affiliation{\FNAL}
\author{C.~Barnes} \affiliation{\Michigan}
\author{G.~Barr} \affiliation{\Oxford}
% ...

\collaboration{The MicroBooNE Collaboration}
\email[]{microboone\_info@fnal.gov}

\date{\today}

\begin{abstract}
We present the first measurement of the single-differential $\nu_e + \bar{\nu}_e$ charged-current inclusive cross sections on argon in electron or positron energy and in electron or positron scattering angle over the full range. Data were collected using the MicroBooNE liquid argon time projection chamber located off-axis from the Fermilab Neutrinos at the Main Injector beam over an exposure of $2.0\times10^{20}$ protons on target. The signal definition includes a 60~MeV threshold on the $\nu_e$ or $\bar{\nu}_e$ energy and a 120~MeV threshold on the electron or positron energy. The measured total and differential cross sections are found to be in agreement with the GENIE, NuWro, and GiBUU neutrino generators.

\end{abstract}

\maketitle
Current and next generation precision neutrino oscillation experiments aim to probe \emph{CP} violation in the lepton sector, the neutrino mass ordering, and physics beyond the Standard Model such as the existence of light sterile neutrinos~\cite{dunecollaboration2020longbaseline, protocollaboration2018hyperkamiokande} by measuring the oscillation of muon neutrinos into electron neutrinos.
Oscillation measurements are particularly sensitive to hard-to-model nuclear effects in the neutrino-nucleus interaction, especially for heavy target nuclei ~\cite{PhysRevLett.123.131801,PhysRevD.98.012004,PhysRevLett.119.082001}. Potentially sizable uncertainties on the $\nu_{e}$/$\nu_{\mu}$ cross section ratio \cite{Day:2012gb,Nikolakopoulos:2019qcr}  reduce the $\nu_{\mu}$'s constraining power.
Only a handful of independent direct measurements of electron-neutrino cross sections exist \cite{PhysRevLett.116.081802,EICHTEN1973281,t2k_nue,Abe2020,PhysRevLett.116.081802}  -- even fewer on argon \cite{Acciarri:2020lhp,PhysRevD.104.052002} -- yet, they are crucial to further understand different flavor neutrino interactions. 

We present a measurement of the $\nu_e + \bar{\nu}_e$ charged current (CC) inclusive cross section on argon at the MicroBooNE experiment. Electrons and positrons are indistinguishable in MicroBooNE and will collectively be referred to as electrons in this paper. The $\nu_e + \bar{\nu}_e$ CC cross section is measured for the first time as a single-differential function of the electron energy in the range 120 MeV to 6 GeV, and as a single-differential function of the electron scattering angle over the full range. 
The contributions from each of the neutrino and antineutrino components are averaged according to their respective fluxes. This is the first demonstration of electron energy reconstruction from $\nu_{e}$ or $\bar{\nu}_{e}$ CC interactions in argon in the $\sim$1 GeV energy range.
The inclusive CC process, in which only the outgoing electron is required to be reconstructed, provides a test of theoretical predictions with minimal dependence on the modeling of the hadronic part of the interaction.

The MicroBooNE detector, which contains 85 tonnes of liquid argon active mass, is located on-axis to the Booster Neutrino Beam (BNB) at Fermilab and $\sim$8$^\circ$ off-axis to the Neutrinos at the Main Injector (NuMI) beam~\cite{ADAMSON2016279}. The NuMI neutrino flux at MicroBooNE contains a $\sim$2\% component of $\nu_{e}$ and $\bar{\nu}_{e}$ with energies ranging from tens of MeV to $\sim$10~GeV at this off-axis angle. For energies above 60~MeV, the $\nu_e$ and $\bar{\nu}_e$ flux is dominated by decays from unfocused kaons at the target. The average $\nu_e$ and $\bar{\nu}_e$ energy is 768~MeV and 961~MeV respectively.

Neutrinos interacting in the MicroBooNE detector create charged particles that traverse a volume of highly pure liquid argon, ionizing the argon and leaving a resulting trail of freed electrons along their paths. The ionization electrons are drifted by an electric field of 273.9 V/cm to a series of three anode wire planes located 2.5 m from the cathode plane, where they induce signals on the wires that are amplified and shaped by front-end electronics immersed in the liquid argon \cite{noise_paper}.  In addition to liberating ionization electrons, the charged particles generate prompt scintillation light as they travel through the medium. The scintillation photons are detected with an array of 32 photomultiplier tubes (PMTs) that are situated behind the anode wire planes \cite{pmt_paper}.

 The NuMI beam operated at medium energy in forward horn current (neutrino) mode for the data used in this analysis. The integrated exposure is 2.0 $\times 10^{20}$ protons on target (POT) after applying data quality criteria for the beam and detector operating conditions.  Two different data streams are used in this analysis: a beam-on data sample collected during the NuMI neutrino spills, and a beam-off data sample acquired in anti-coincidence with the neutrino beam. The beam-off data sample is used to model the cosmic ray (CR) backgrounds -- a essential task given MicroBooNE's location on the surface.

A GEANT4-based \cite{geant} simulation of the NuMI beamline is used for generating the neutrino flux prediction. The simulation models the interactions of protons on the NuMI graphite target and the subsequent particle cascade, decay chain, and reinteractions. Hadron production is constrained using data from the NA49 experiment \cite{Alt2007} and other applicable measurements with the Package to Predict the FluX (PPFX) software package~\cite{Aliaga:2016oaz}. 

The simulation and reconstruction of the events are performed using the LArSoft framework~\cite{larsoft}. Neutrino interactions in the MicroBooNE detector are simulated using the GENIE v3.0.6 event generator~\cite{GENIE} where the CC quasi-elastic (QE) and CC meson exchange current (MEC) neutrino interaction modes are tuned \footnote{Denoted as GENIE v3.0.6 ($\mu$B tune) in this paper.} to $\nu_\mu$ CC 0$\pi$ data from T2K~\cite{PhysRevD.93.112012,themicroboonecollaboration2021new}. GENIE generates all final state particles associated with the primary neutrino interaction along with the transport and rescattering of these final state particles through the target nucleus.

Particle propagation in the MicroBooNE simulation is based on GEANT4. The energy depositions from charged particles are processed with a dedicated series of algorithms, starting with simulation of long-range electronic signals induced on the TPC anode wires by drifting ionization electrons~\cite{signal_process_part1,signal_process_part2}. Optical signals of the energy depositions on the PMTs are also simulated. 

The simulated neutrino interactions are overlaid with beam-off data which provides a data-driven model for CRs crossing the detector volume within the readout window of neutrino events. Events from data and simulation are processed and calibrated according to the standard MicroBooNE chain described in Ref. ~\cite{noise_paper,signal_process_part1,signal_process_part2,MicroBooNE_calibration,showerEn}, and reconstructed with  the Pandora pattern recognition framework \cite{pandora}.

To select signal candidate events, this analysis combines information from the TPC event topology --- number of final state particles, vertex candidate vertical position, average particle direction, and activity near the vertex --- with information from the optical system \cite{VanDePontseele:2020tqz}. Requiring the containment of the reconstructed neutrino vertex and a high number of associated hits within a fiducialized portion of the TPC abates out-of-TPC and CR backgrounds. Only events with at least one reconstructed shower are selected. Showers are identified using the track-score variable from Pandora~\cite{VanDePontseele:2020tqz}. To remove background events such as $\nu_\mu$ CC $\pi^0$ and NC $\pi^0$, selections on the deposited energy per unit length ($dE/dx$) at the beginning of the shower, the distance to the neutrino vertex, and the transverse profile are applied on the shower with the highest number of hits. 

The cross section is presented as a function of electron energy and angle. The angle, $\beta$, represents the electron's deflection from the neutrino direction. When reconstructing $\beta$, we assume all neutrinos originated from the beam target. The true direction of $\sim$95\% of the selected simulated electron neutrinos and antineutrinos is within 3 deg from this assumption. The resolution in $\cos\beta$ ranges from 0.01 to 0.05. The shower is rarely misreconstructed with the opposite direction (0.2\% of selected events). The electron energy resolution is described by a Gaussian distribution with exponential tail which presents a low-sided bias ranging from (3-14)\% and standard deviation ranging from (15-30)\%.

The final selected sample contains 243 events. The selection has an average $\nu_e$ + $\bar{\nu}_e$ efficiency of 21\% and an individual efficiency of 20\% for $\nu_e$ and 24\% for $\bar{\nu}_e$. The higher efficiency for $\bar{\nu}_e$ is due to the higher mean energy of these neutrinos where the analysis is more efficient. The final purity of the analysis is 72\%. The selected signal sample is predicted to be 48\% CC QE, 28\% CC resonant (RES), 17\% CC MEC and 7\% CC deep-inelastic scattering (DIS) according to GENIE v3.0.6 ($\mu$B tune).

\begin{figure*}[]
  \centering
  \subfigure[]{\includegraphics[width=0.45\textwidth]{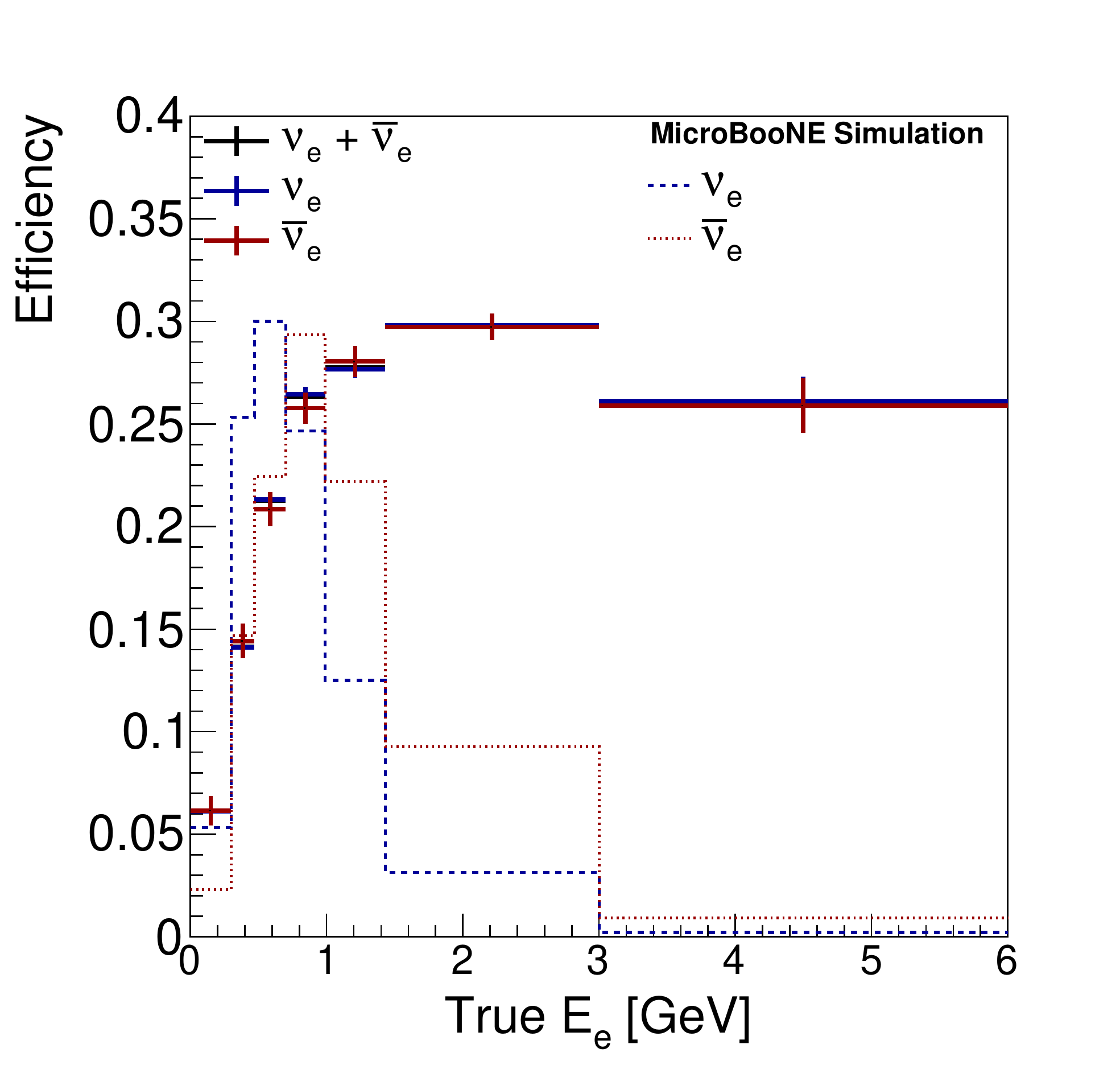}}\quad\quad\quad
  \subfigure[]{\includegraphics[width=0.45\textwidth]{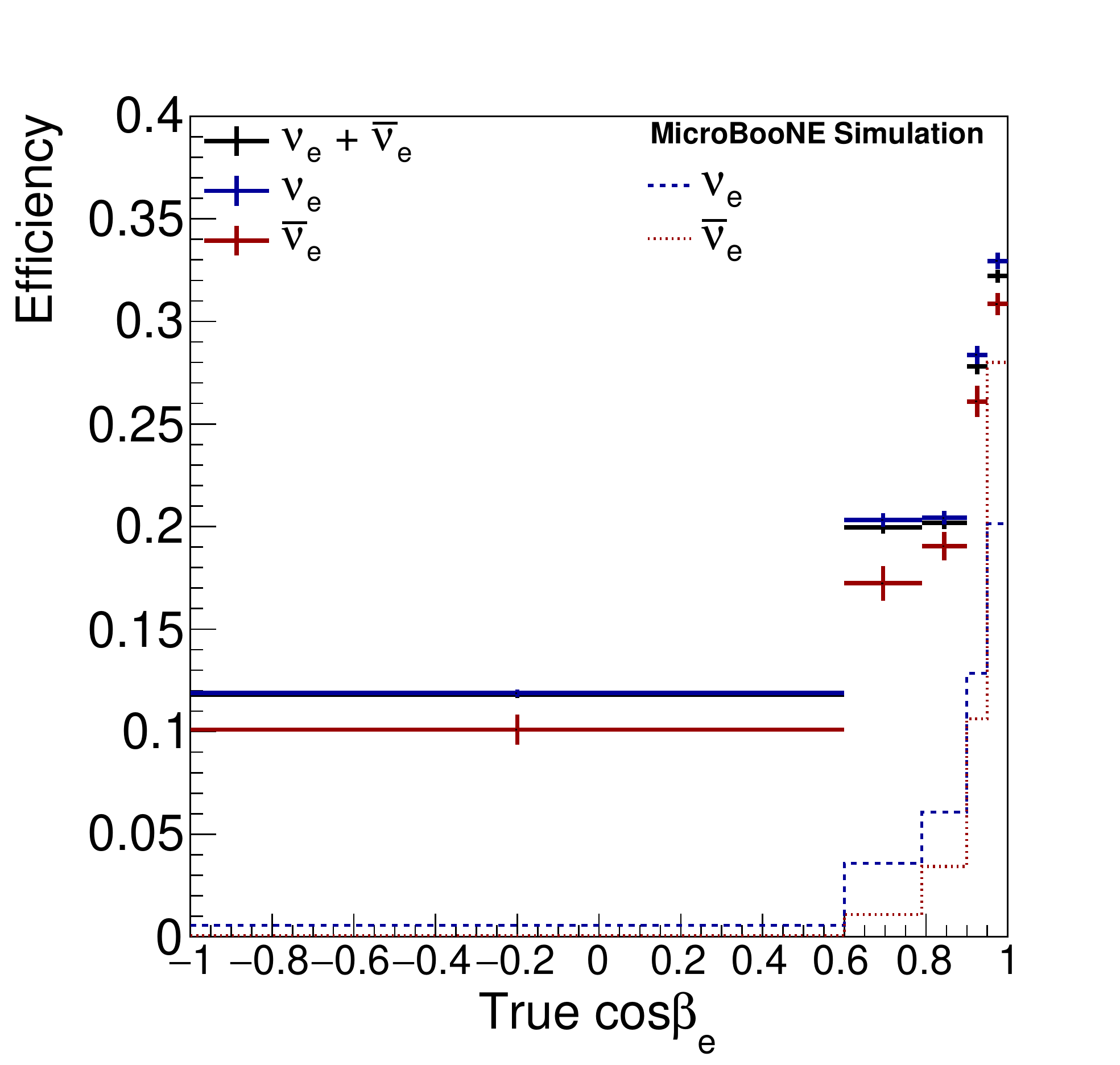}}
  \caption{Simulated efficiency broken down by $\nu_e$, $\bar{\nu}_e$ and $\nu_e + \bar{\nu}_e$ as a function of the electron (a) energy and (b) angle. The error bars include statistical uncertainties only. The distributions with dashed lines show the area normalized predicted event distributions before selection.}
  \label{fig:eff}
\end{figure*}

Figure~\ref{fig:eff} shows the efficiency as a function of the kinematic variables. The efficiency decreases towards lower energies because the electrons stop producing sizable showers which are the key feature recognized by the selection algorithms. At higher energies, above 3~GeV, DIS interactions become the primary channel. For the purposes of this analysis, the Pandora reconstruction algorithm was not tuned on high multiplicity events. The many particles resulting from these interactions can hinder the electron-induced shower reconstruction thus lowering the efficiency.

The main backgrounds in this analysis are (i) CRs in time with the beam spill (estimated to be 8.3\% of all selected events), (ii) neutral current interactions containing a $\pi^0$  (7.0\%), and (iii) charged current $\nu_{\mu}$ or $\bar{\nu}_{\mu}$ interactions with a $\pi^0$ in the final state (4.2\%). The remaining backgrounds include neutrino-induced activity outside the fiducial volume and NC interactions without $\pi^0$ in the final state. Only events with a true electron energy above 120~MeV are considered signal. In addition, a 60~MeV threshold for the $\nu_{e}$ or $\bar{\nu}_{e}$ energy is used in the lower bound in the integral that calculates $\Phi$ in Eqn. \ref{eq:xsec_differential}. Selected signal events below these thresholds form a negligible background. The CR backgrounds are modeled using beam-off data. All other backgrounds are estimated from the simulation. The accuracy of the detector modeling has been verified by studying selected event distributions using quantities not affected by the neutrino interaction physics, for example, the neutrino interaction locations in the detector.

We report the differential cross section as a function of true kinematic variables using the Wiener single value decomposition (Wiener-SVD) unfolding technique~\cite{Tang:2017rob}. This method corrects a measured  differential event rate, defined in Eqn.~\ref{eq:xsec_differential}, for inefficiency and finite resolution. The correction is performed by minimizing a $\chi^2$ score that compares data to a prediction and includes a regularization term. The degree of regularization is determined from a Wiener filter that is used to minimize the mean square error between the variance and bias of the result. In addition to the measured event rate, the 
inputs to the method are a covariance matrix calculated from 
simulation (which approximately describes the statistical and systematic uncertainties on the measurement), and a response matrix that describes the detector smearing and efficiency. 
The Wiener-SVD method produces an \textit{unfolded} differential cross section in true kinematics, a covariance matrix describing the total uncertainty on this cross section, and an additional smearing matrix, $A_{c}$, which contains information about the regularization and bias of the measurement. The matrix $A_{c}$ is applied to a true cross section prediction when comparing to the unfolded data.

The flux-averaged, differential event rate as a function of a variable $x$ is defined as,
\begin{equation}
\label{eq:xsec_differential}
\left ( \frac{dR}{dx} \right )_{i} = \frac{N_{i} - B_{i}}{T \times \Phi \times \Delta x_i},
\end{equation}
where $N_i$, $B_i$, and $\Delta x_i$ are the number of selected events, the expected background events, and bin width in bin $i$ respectively, $T$ is the number of target nucleons, and $\Phi$ is the total POT-scaled NuMI $\nu_e+\bar{\nu}_e$ flux (integrated from 60~MeV). The flux corresponding to $2.0\times10^{20}$ POT is 
$1.845 \times 10^{11}/$cm$^2$ which has a mean energy of 837~MeV. 

The statistical and systematic uncertainties on $dR/dx$ are encoded in the total covariance matrix, $E_{ij} = E_{ij}^{\textrm{stat}} + E_{ij}^{\textrm{syst}}$, where $E_{ij}^{\textrm{stat}}$ is a diagonal covariance matrix including the statistical uncertainties and $E_{ij}^{\textrm{syst}}$ is a covariance matrix including the total systematic uncertainties.

The PPFX package is used to assess the hadron production uncertainties on the neutrino flux prediction by reweighting the nominal simulation. This consists of creating a number of replica simulations, each one called a ``universe". A set of weights is produced by sampling the hadron production parameters within their respective uncertainties. The procedure accounts for uncertainties in the flux shape addressing issues raised in Ref.~\cite{PhysRevD.102.113012}. A similar method is used for evaluating the uncertainties on the cross section model but sampling the parameters used in GENIE within their uncertainties \cite{GENIE, GENIE_reweighting}. This technique reweights all model parameters simultaneously, enabling a correct treatment of correlations among the parameters. A total of $s$ such universes are used to construct a covariance matrix, \begin{equation}
\label{eq:multisim}
E_{ij} = \frac{1}{s} \sum_{n = 1}^{s} (R_i^n - R_i^\text{cv})(R_j^n - R_j^\text{cv}),
\end{equation} where $R_i^\text{cv}$($R_j^\text{cv}$) and $R^n_i$($R^n_j$) are the flux-averaged event rates for the central value and  systematic universe $s$ in a measured bin $i$($j$) respectively.

\begin{figure*}[]
  \centering
  \subfigure[]{\includegraphics[width=0.45\textwidth]{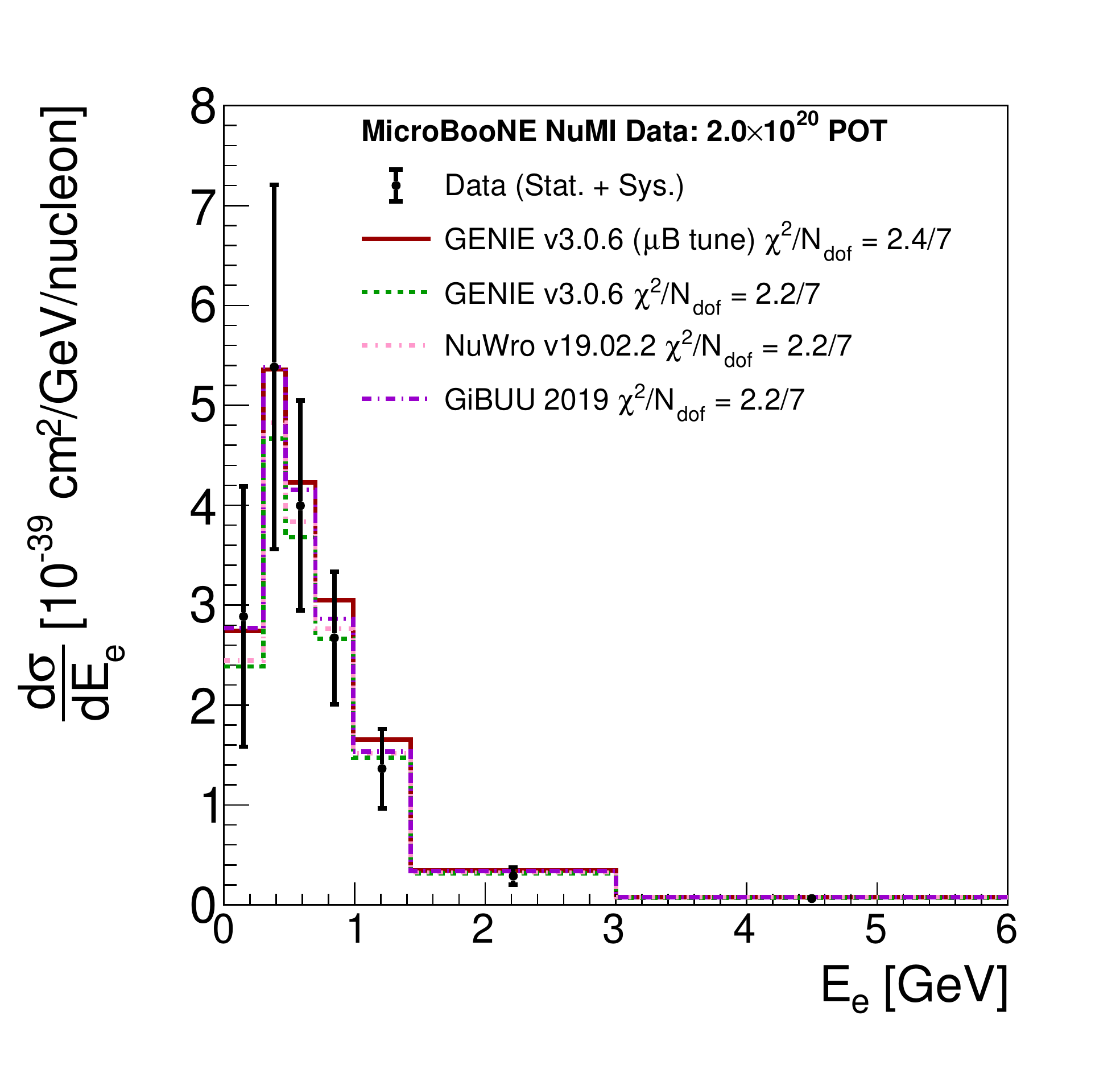}}\quad\quad\quad
  \subfigure[]{\includegraphics[width=0.45\textwidth]{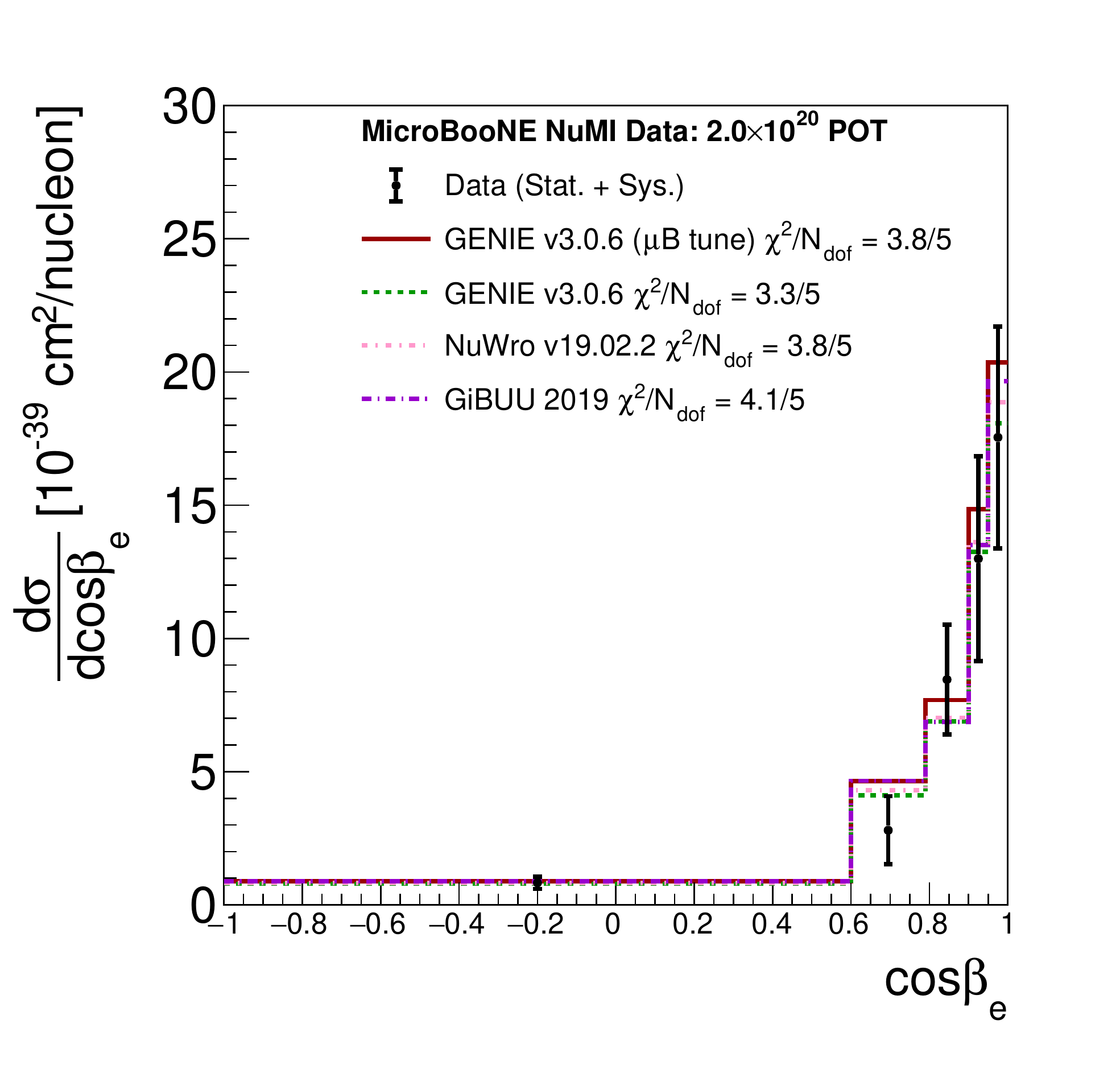}}
  \caption{Unfolded differential cross section as a function of the electron (a) energy and (b) angle. The data cross section is compared to GENIE v3.0.6 ($\mu$B tune)(red), GENIE v3.0.6 (green), NuWro v19.02.2 (pink) and GiBUU 2019 (purple), and is in agreement with all predictions.}
  \label{fig:unfolded}
\end{figure*}

A different method is followed for systematic uncertainties associated with the detector model, the NuMI beamline geometry modeling, and additional cross section modeling not encapsulated by the GENIE multi-parameter reweighting. These systematic uncertainties are obtained by using single-parameter variation, in which only one parameter at a time is changed by its estimated $1\sigma$ uncertainty. For $s$ parameters, the covariance matrix is given by,
\begin{equation}
\label{eq:unisim}
E_{ij} = \sum_{m = 1}^{s} (R_i^m - R_i^\text{cv})(R_j^m - R_j^\text{cv}).
\end{equation}

\noindent A summary of all uncertainties on the total data cross section is shown in Table \ref{tab:total_xsec_uncertainties}. 

\begin{table}[h]
\caption{\label{tab:total_xsec_uncertainties}Contributions to the total data cross section measurement uncertainty.}
\begin{ruledtabular}
\begin{tabular}{lc}
Source of Uncertainty        &  Relative Uncertainty [\%] \\ \hline
\colrule
Beam Flux                &  17.4 \\
Detector                 &  6.8  \\
Cross Section            &  5.8  \\
POT Counting             &  2.0  \\
Out-of-Cryostat          &  1.8  \\
Proton/Pion Reinteractions    &  1.2  \\
Beam-off Normalization   &  0.1  \\
\colrule
Total Systematic Uncertainty         &  19.8 \\
\colrule
MC Statistics            &  0.8  \\
Data Statistics          &  10.0 \\
\colrule
Total Uncertainty        &  22.2 \\ 
\end{tabular}
\end{ruledtabular}
\end{table}

For the differential cross section measurement, statistical uncertainties in each bin are the largest source of uncertainty. The most significant contributions to the systematic uncertainty are the hadron production flux uncertainties, especially from hadrons produced by secondary nucleons which interact with non-carbon-based materials and meson interactions not covered by any existing hadron production data. The hadron production uncertainties are largest ($\sim$30\%) for low energies ($<$300~MeV) and range from (15-20)\% near the peak of the event distribution. 

The second-largest source of uncertainty comes from a combination of detector-based uncertainties in light yield, ionization electron  recombination model, space-charge effect~\cite{Abratenko_2020bbx}, and waveform deconvolution~\cite{microboonecollaboration2021novel}. These uncertainties are the most significant uncorrelated contributions to the total covariance matrix but result in subdominant contributions compared to the statistical uncertainties per bin. Other sub-leading uncertainties include uncertainties on the cross section modeling, the modeling of proton and pion transportation in argon, the total POT recorded by the NuMI beamline monitors, out-of-cryostat modeling, and normalization of the beam-off to beam-on data.

The unfolded differential cross section in electron energy and angle is presented in Fig.~\ref{fig:unfolded} and is compared with GENIE v3.0.6 ($\mu$B tune), NuWro v19.02.2, GiBUU 2019, and an untuned version of GENIE v3.0.6. All generator predictions are smeared with the matrix $A_c$. The models used in GENIE v3.0.6 \footnote{The GENIE \texttt{G18\_10a\_02\_11a} comprehensive model configuration is used.} include a Local Fermi Gas (LFG) nuclear~\cite{CARRASCO1992445} model and a Nieves CC QE~\cite{PhysRevD.85.113008} model. Coulomb corrections for the outgoing lepton~\cite{PhysRevC.57.2004} and Random Phase Approximation corrections (RPA)~\cite{PhysRevC.70.055503} are applied. A Nieves model is used for MEC~\cite{schwehr2017genie}, a Kuzmin-Lubushkin-Naumov~\cite{KLN} and Berger-Seghal~\cite{Minibooneq2,PhysRevD.76.113004} model is used for RES, and Berger-Seghal is used for Coherent (COH)~\cite{PhysRevD.79.053003} interactions. Final State Interactions (FSI) are modeled using an empirical hA2018 model~\cite{PhysRevC.23.2173}. NuWro uses similar models to GENIE which include a LFG nuclear model with a binding energy derived from a potential. A Llewellyn-Smith~\cite{LLEWELLYNSMITH1972261} QE model is used with RPA corrections that are implemented with a different treatment to the Nieves model used within GENIE. To model multi-nucleon interactions, a transverse enhancement model~\cite{Eur.Phy.JC71.1726} is used. Resonant interactions use an Adler-Rarita-Schwinger model which calculates $\Delta$(1232) resonance explicitly and includes a smooth transition to DIS at 1.6~GeV~\cite{Sobczyk:2004va}. DIS interactions use a Bodek-Yang~\cite{Bodek_2003,Yang_2000} model and a Berger-Sehgal~\cite{PhysRevD.79.053003} model for COH interactions. For FSI, a Salcedo-Oset model is used for pions~\cite{SALCEDO1988557} and nucleon-medium corrections are used for nucleons~\cite{PhysRevC.45.791}. GiBUU 2019 \cite{Buss_2012} includes consistent nuclear medium corrections throughout and uses a LFG nuclear model~\cite{CARRASCO1992445}. An empirical MEC model is used \cite{PhysRevC.94.035502}, and final state particles are propagated according to the Boltzmann-Uehling Uhlenbeck transport equations.

 The $\chi^2$ per degree of freedom (d.o.f.) data comparison for each generator takes into account the total covariance matrix including the off-diagonal elements. For the electron energy the values of $\chi^2$/d.o.f. range from 2.2/7 to 2.4/7 (GENIE v3.0.6, $\mu$B tune), while they range from 3.3/5 (GENIE v3.0.6) to 4.1/5 (GiBUU) for the cos$\beta$, see Fig.~\ref{fig:unfolded} a) and b). The measurement is therefore in agreement with all considered models for both electron energy and angle. The value of the $\chi^2$/d.o.f. reported for each model is predominantly driven by the data statistical uncertainty with the systematic uncertainty contributing to its small value.

This measurement is the first test of multiple neutrino event generators against electron neutrino and antineutrino differential scattering data on argon. It is sensitive to CC QE, CC RES, CC MEC and CC DIS scattering with full angular coverage and for electron energies ranging from 120~MeV - 6~GeV. Supplemental materials include cross section values, efficiencies, purity, flux, additional smearing matrices, uncertainties in each bin, and unfolded covariance matrices. 

Additionally, the flux-averaged total data cross section is calculated as $(4.90\pm\!0.49~\textrm{(stat.)}\pm\!0.97~\textrm{(sys.))}\!\times\!10^{-39}~\textrm{cm}^{2}/~\textrm{nucleon}$. This agrees with the GENIE v3.0.6 ($\mu$B tune), GENIE, NuWro and GiBUU predictions within uncertainties. Moreover, the total cross section agrees with MicroBooNE's previous measurement~\cite{microboonecollaboration2021measurement} within 3\% (when adjusted for the different signal definitions) while reducing the uncertainty by almost a factor of two.

In summary, this letter presents the first single-differential electron neutrino and antineutrino cross section on argon as a function of the electron energy and scattering angle over the full range. The measurement is compared to several generators including GENIE v3.0.6 ($\mu$B tune), GENIE v3.0.6, NuWro v19.02.2, and GiBUU 2019, and is in agreement for all predictions. This measurement provides an excellent test and validation of neutrino-nucleus generators on argon and will be valuable for the short-baseline programs such as SBN and searches for \emph{CP} violation with long-baseline experiments such as DUNE~\cite{dunecollaboration2020longbaseline} for which electron neutrino interactions on argon are the primary signal channel.

This document was prepared by the MicroBooNE collaboration using the
resources of the Fermi National Accelerator Laboratory (Fermilab), a
U.S. Department of Energy, Office of Science, HEP User Facility.
Fermilab is managed by Fermi Research Alliance, LLC (FRA), acting
under Contract No. DE-AC02-07CH11359.  MicroBooNE is supported by the
following: the U.S. Department of Energy, Office of Science, Offices
of High Energy Physics and Nuclear Physics; the U.S. National Science
Foundation; the Swiss National Science Foundation; the Science and
Technology Facilities Council (STFC), part of the United Kingdom Research and Innovation; and The Royal Society (United Kingdom).  Additional support for the laser calibration system and CR tagger was provided by the Albert Einstein Center for Fundamental Physics, Bern, Switzerland.
\bibliographystyle{apsrev4-1}
\bibliography{Bibliography}
\end{document}